\documentclass[aps,prd,groupedaddress,superscriptaddress,12pt]{revtex4}

\usepackage{amsmath,amssymb,bm}
\usepackage{graphicx}
\usepackage{epstopdf}
\usepackage{amsfonts}
\usepackage{amssymb}
\usepackage{amsbsy}
\usepackage{amsmath}
\usepackage{latexsym}
\usepackage{sansmath}
\usepackage{bm}
\usepackage{color}
\usepackage{comment}
\usepackage{csquotes}								
\usepackage{physics}
\usepackage{eufrak}
\usepackage{accents}
\usepackage[colorlinks=true,breaklinks]{hyperref}

\newcommand{\eq}[1]{Eq.\,(\ref{#1})}

\def\be{\begin{equation}}
\def\ee{\end{equation}}
\def\bea {\begin{eqnarray}}
\def\eea {\end{eqnarray}}
\def\nn {\nonumber}
\def \p {\partial}

\def \l {\left}
\def \r {\right}

\newcommand{\Hd}{\mathcal{H}_D}

\newcommand{\di}{\partial}

\newcommand{\smh}[1]{{\color{red}{[{\bf SMH}: #1]}}}    

\newcommand{\ma}[1]{{\color{blue}{[{\bf MA}: #1]}}}
\newcommand{\Hp}{\mathcal{H}_{\text{p}}}

\begin{document}

\title{The Universe as an oscillator}
\author{Masooma Ali} \email{masooma.ali@unb.ca} 
\affiliation{Department of Mathematics and Statistics, University of New Brunswick, Fredericton, NB, Canada E3B 5A3}
\author{Syed Moeez Hassan} \email{shassan@unb.ca} 
\affiliation{Department of Mathematics and Statistics, University of New Brunswick, Fredericton, NB, Canada E3B 5A3}
\author{Viqar Husain} \email{vhusain@unb.ca} 
\affiliation{Department of Mathematics and Statistics, University of New Brunswick, Fredericton, NB, Canada E3B 5A3}
\affiliation{Perimeter Institute for Theoretical Physics, 1 Caroline St N, ON, Canada}

\bigskip

\begin{abstract}

We apply the idea of using a matter time gauge in quantum gravity to quantum cosmology. For the Friedmann-Lemaitre-Robertson-Walker (FLRW) Universe  with dust and cosmological constant $\Lambda$, we show that the dynamics maps exactly to the simple harmonic oscillator in the dust-time gauge. For $\Lambda >0$ the oscillator frequency is imaginary, for $\Lambda<0$ it is real, and for $\Lambda=0$ the Universe is a free particle. This  result provides (i) a simple demonstration of non-perturbative singularity avoidance in quantum gravity for all $\Lambda$, (ii) an exact Lorentzian Hartle-Hawking wave function, and (iii) gives the present age of the Universe as the characteristic decay time of the propagator.   
\end{abstract}

\maketitle

\section{Introduction}

The application of quantum theory to gravity is pursued using a number of different approaches (see e.g. \cite{Duff:2012sqa} for a recent survey). These can be broadly divided into two -- those that are ``background dependent" and those that are not \cite{Isham:1993ji}. The term refers to what structures in the classical theory are to be held fixed in the passage to  quantum theory. The canonical quantization approach formulated by DeWitt \cite{DeWitt:1967yk} is considered to be the defining case of a background independent approach to quantum gravity; this is also the paper where the very first quantization of the FRLW model was described.

The canonical quantization program naturally divides into two distinct approaches. These are referred to as (i) Dirac quantization, where the Hamiltonian constraint is imposed as an operator condition on wave function(al)s, and (ii) reduced phase space quantization, where time and spatial coordinate gauges are fixed in the classical theory before proceeding to quantization. It is the former that leads to the Wheeler-DeWitt equation. Solutions in either case are referred to as ``wave functions of the Universe."

In its more recent incarnations, the Dirac quantization condition is approached via a path integral as in the Hartle-Hawking method \cite{Hartle:1983ai}, or by imposing the condition directly as in Loop Quantum Gravity \cite{Ashtekar:2004eh,Thiemann:2007pyv}.  In reduced phase space quantization, a phase space variable is first selected as a clock. Its conjugate variable provides the physical non-vanishing Hamiltonian.  Quantization then proceeds as in conventional quantum theory with a time-dependent Schrodinger equation (or path integral). This division has led to much debate about the role of time in quantum gravity at both the philosophical and physical levels, and questions about the equivalence of the two methods \cite{Hartle:1984ut,Schleich:1990gd,  Giesel:2017mfc}.   

Because of the difficulty in solving the Wheeler-deWitt equation in the former case and the time-dependent Schrodinger equation in the latter case, nearly all concrete calculations are restricted to either homogeneous cosmological models or to inhomogeneous perturbations of these models. Examples of early work  on such models include Refs.  \cite{Misner:1969hg, Blyth:1975is}. The most recent works   are in the framework of Loop Quantum Cosmology (LQC) \cite{Agullo:2016tjh}, and a revisit of the Hartle-Hawking prescription via a Lorentzian path integral \cite{Feldbrugge:2017kzv,DiazDorronsoro:2017hti}.

In this note we revisit the flat, homogeneous and isotropic  cosmology with dust and a cosmological constant $\Lambda$. This remains the  typical  model to consider since current observations suggest  that our Universe is modelled well by an FLRW cosmology with zero spatial curvature and a very small positive cosmological constant $\Lambda \sim 3 \times 10^{-122} l_P^{-2}$ \cite{Ade:2015xua}.  We study the  model using the reduced phase method in the dust time gauge  \cite{Brown:1994py,Husain:2011tk,Husain:2011tm,Giesel:2012rb,Ali:2015ftw}; a recent study via Dirac-Wheeler-deWitt  quantization appears in \cite{Maeda:2015fna}.  In the context of matter time gauges,  there are also several studies using scalar field time in quantum gravity and cosmology; a representative selection is \cite{Rovelli:1993bm,Feinberg:1995tf,Nakonieczna:2015dza, Assanioussi:2017tql}.  

 In the Arnowitt-Deser- Misner (ADM) canonical formalism, we show that dust time gauge leads to a surprising result: the corresponding  physical Hamiltonian, after a canonical transformation,  becomes  exactly that of a simple harmonic oscillator;  the oscillator's frequency is determined by $\sqrt{\Lambda}$.  The corresponding quantum theory is therefore immediate. 
 
 For $\Lambda<0$ the potential is that of the usual oscillator, whereas  for $\Lambda >0$  it is the inverted oscillator.  The former case describes Universes either as stationary states, or as wave packets that expand and contract ad-infinitum. The latter case has only scattering solutions that give Universes with a single bounce. Depending on the choice of canonical parametrization, the oscillator is either on the half or the full line. All cases gives singularity avoidance, and for all choices of self-adjoint extensions of the Hamiltonian.  Our work also exhibits one of the situations where Dirac and reduced phase space quantization give similar results for a particular choice of operator ordering in the Wheeler-Dewitt equation.
 
 We begin by reviewing the general formalism for the dust time gauge, followed by its application to cosmology in the following sections.

\section{Dust time gauge}

 The model we consider is general relativity coupled to a pressureless dust field $T$. The  action
\be
S= \int d^4x ~  \sqrt{-g}R  -   \int d^4x ~  \dfrac{1}{2} M \sqrt{-g} \left( g^{ab} \p_aT \p_bT + 1 \right)
\ee
where $g_{ab}$ is the 4-metric, $R$ is the 4-Ricci Scalar, and $M$ is the dust energy density, leads to the canonical ADM action 
\be
 S= \int d^3x\  dt  \ \left(\pi^{ab} \dot{q}_{ab}  + p_T \dot{T} - N{\cal H} - N^a {\cal C}_a   \right) , 
\ee
where
\begin{subequations} 
\bea
{\cal H} &\equiv&  {\cal H}_G +  \Hd \nn\\
              &=& 
   \frac{1}{\sqrt{q}} (\pi^{ab}\pi_{ab} - \frac{1}{2} \pi^2 ) +   \sqrt{q} (\Lambda - {}^{(3)} \!R) \nn\\
   && + \  \text{sgn}(M)\  p_T \sqrt{1+q^{ab}\p_aT \p_bT},  \\	 
{\cal C}_a &\equiv&  -D_b \pi^b_{\ a}  + p_T \di_a T,
\eea
\end{subequations}
$q_{ab}$ is the 3-metric, $\pi^{ab}$ is its conjugate momentum, $p_T$ is the dust conjugate momentum, $N$ is the lapse, $N^a$ is the shift, and the metric is of the ADM form
 \be
ds^2 = -N^2dt^2 + (dx^a + N^a dt)(dx^b + N^b dt)q_{ab}. \label{admm}
\ee  

We define the canonical dust time gauge  by 
 \be
 T=\epsilon t
 \ee
with $\epsilon= \pm 1$. The requirement  that the gauge be preserved in time gives 
\be
 \dot{T} = \epsilon = \Big\{ T, \int d^3x\   N {\cal H} \Big\}\Big|_{T=t} = \text{sgn}(M) N. \label{gauge-pres}
\ee 
The physical Hamiltonian $\Hp$ is obtained by substituting the gauge into the dust symplectic term in the canonical action, which identifies 
$\Hp \equiv -\epsilon p_T$. Solving the Hamiltonian constraint 
\be
{\cal H}_G + \text{sgn}(M) p_T = 0 
\ee
then identifies the physical Hamiltonian 
\be
\Hp = -\epsilon p_T  =   \epsilon\ \text{sgn}(M)   {\cal H}_G   = N   {\cal H}_G  \ ,
\ee
using  (\ref{gauge-pres}) for the last equality.  It is also useful to note, using $p_T = \sqrt{q}\ \dot{T}M/N$ and (\ref{gauge-pres}), the relation
\be
  p_T = \epsilon  \sqrt{q}\  \frac{M}{N} = \epsilon \ \sqrt{q}\  \frac{\text{sgn}(M)}{N}\ |M| = \sqrt{q}\  |M| \ \ .    
\ee
which shows that  $p_T > 0$  for $M\ne 0$, and 
\be
\Hp = -\epsilon \sqrt{q}\  |M| = N   {\cal H}_G  .  
\ee  
Thus the requirement that the dust Hamiltonian satisfy ${\cal H}_D = \text{sgn}(M)p_T\ge 0$ implies $\text{sgn}(M)=+1$, since $p_T= \sqrt{q}\ |M| \ge 0$. This means that the dust field satisfies the weak energy condition. With this choice (\ref{gauge-pres}) gives $N=\epsilon$.  In the following we make the choice $N=\epsilon = -1$ which gives the manifestly positive physical Hamiltonian density
\be
\Hp = \sqrt{q}\ |M| = -  {\cal H}_G   \ge 0. \label{Hp2}
\ee

 \subsection{Application to cosmology}
 
 Let us now consider the reduction of the dust-time gauge theory to homogeneous and isotropic cosmology This is obtained by setting
 \bea
  q_{ab} &=& a^2(t) e_{ab}\nn \\
  \pi^{ab} &=& \frac{p_a(t)}{6a(t)}  e^{ab},
 \eea
 where $e_{ab} = \text{diag}(1,1,1)$ is a fiducial flat metric. The reduced phase space coordinates are $(a, p_a)$, and we take $a \in (0,\infty)$ and $p_a \in \mathbb{R}$ as the definition of this parametrization (since we must have $\text{det}(q_{ab}) = a^3 > 0$).  
 
 The physical Hamiltonian  (\ref{Hp2}) for the flat case then becomes
 \be
 \Hp = \frac{p_a^2}{24a} - \Lambda a^3.
 \ee
 To briefly recap, this FRW model started with a four-dimensional phase space, that of the dust field and the scale factor. After fixing the time gauge and solving the Hamiltonian constraint, the reduced phase space becomes two-dimensional, with canonical coordinates $(a,p_a)$. This is unlike the vacuum deSitter model  (see e.g.\cite{Halliwell:1988ik} ),  which actually has no physical degrees of freedom;  the physical meaning of ``wave functions of the Universe" without additional degrees of freedom is therefore unclear.    
  
Let us now note the canonical transformation 
\be
p= \frac{p_a}{\sqrt{12a}},\ \ \ \  x = \frac{4}{\sqrt{3}} a^{3/2} \label{canon}
\ee
and the rescaling $\Lambda \longrightarrow 4\Lambda/\sqrt{3}$ transforms the Hamiltonian to 
\be
\Hp = \frac{1}{2}\left({p^2} - \Lambda x^2 \right).  \label{Hpfrw}
\ee
 There are thus three cases of interest: $\Lambda=0$ is a free particle, $\Lambda<0$ is the oscillator and $\Lambda>0$ is the inverted oscillator.

\section{Quantization and wave functions of the Universe}

This section consists of two parts where we describe quantization in the dust time gauge for two choices of the configuration space. These lead to quantum theories on either the half-line or the full line. In the former case there is a one parameter family of self-adjoint extensions of the physical Hamiltonian.  

\subsection{Quantization on the half-line}

The classical theory is on the half-line, $x\in (0,\infty)$, so the obvious choice for the Hilbert space is  $L^2(\mathbb{R}^+,dx)$. In this space it is known that  Hamiltonians of the form $p^2 + V(x)$  have self-adjoint extensions. Specifically, it is readily checked that the physical Hamiltonian (\ref{Hpfrw}) is symmetric in the usual representation $\hat{p} \rightarrow -i \partial_x$, i.e. that $(\psi, \widehat{\Hp} \phi) = (\widehat{\Hp} \psi, \phi)$, provided $\displaystyle \lim_{x\rightarrow \infty} \phi=0$ and 
\be
\lim_{x\rightarrow 0}\left[\psi^* \phi' - \phi  \psi^{* \prime} \right] = 0.
\ee
This gives the boundary condition $\phi'(0) = \alpha \phi(0), \  \alpha \in \mathbb{R} $. Thus there is a one-parameter ($\alpha$) family of self-adjoint extensions of $\widehat{\Hp}$ on the half-line, so  the Hilbert space is the subspace  specified by 
\be
\mathbb{H}_\alpha =  \left\{ \phi \in {\cal L}^2(\mathbb{R}^+,dx) \Big|  \lim_{x\rightarrow 0} (\ln \phi) '= \alpha \in  \mathbb{R} \right\}.
\ee
We are interested in solving the time-dependent Schrodinger equation,
\be
i \frac{\partial }{\partial t} \phi(x,t) = -\frac{1}{2} \frac{\partial^2}{\partial x^2} \phi(x,t) - \frac{1}{2} \Lambda x^2 \phi(x,t),
\ee
with the boundary condition mentioned above. (In this equation all variables are dimensionless, or equivalently, written in Planck units.)

\noindent \underbar{$\Lambda=0$}:  There are two types of elementary solutions. The first are the ingoing and outgoing waves of fixed energy   (in the dust time gauge), and satisfying the above boundary condition, 
\be
\phi_{\alpha k}(x,t) = e^{-ik^2t/2} \left[e^{ikx} - \left(\frac{\alpha-ik}{\alpha+ik}\right)  e^{-ikx} \right]
\ee
 Normalizable wave functions are constructed in the usual manner as 
 \be
 \psi_\alpha(x,t) =   \int_{-\infty}^\infty dk \ f(k) \phi_{\alpha k}(x,t)
 \ee
All such solutions describe Universes with singularity avoidance and a bounce at the origin with a phase shift given by 
$\alpha$. 
 
The second type of solution is a bound state,  
\be
\phi(x,t) = e^{i\kappa^2t/2}\ e^{-\kappa x}, \ \ \ \kappa>0
\ee
This corresponds to $\alpha = -\kappa$, a choice permitted by the boundary conditions. The Universe this describes is ruled out by experiment,  since $ \langle a^{3/2}\rangle  \sim \langle x \rangle  = (2\kappa)^{-1}$ which has  the interpretation of an emergent flat spacetime from the expectation value of the metric.  
 \medskip
 
\noindent \underbar{$\Lambda <0$}:  This is the oscillator on the half-line with the  boundary condition, $\psi'(0) - \alpha \psi(0)=0$. With $\Lambda = -1/l^2$ and $\zeta=t/l$, the propagator on $\mathbb{R}$ is a basic result, 
\bea
K(x,\zeta; x',0) &=& \sqrt{\frac{1}{2\pi i l \sin{\zeta}}} \nn\\
  &&\times  \exp\left\{ \frac{i[(x^2 +x'^2)\cos \zeta -2xx' ]}{2 l \sin \zeta}    \right\}.\label{prop}
\eea
For the half-line problem at hand, given  initial data  $\psi(x,0)=f(x)$ for $x>0$, the solution with the required boundary condition at $x=0$ may be obtained by extending the given  initial data $f(x)$ on $\mathbb{R}^+$ to  the region $x<0$, such that  
\be
f'(x)-\alpha f(x) = - \left( f'(-x) - \alpha f(-x)\right) \label{f}, \ \ \  x<0,
\ee
i.e. imposing antisymmetry on the boundary condition function. Solving this equation gives the required extension 
\bea
f_L(x) & \equiv& e^{\alpha x}\int_x^0  du\  e^{-\alpha u} \left[ f'(-u) - \alpha f(-u) \right] \nn\\
&& +\ e^{\alpha x} f(0), \ \ \ x<0,  
\eea
where the integration constant is chosen such that $f_L(0) = f(0)$. 

\begin{figure}
\includegraphics[width=\linewidth]{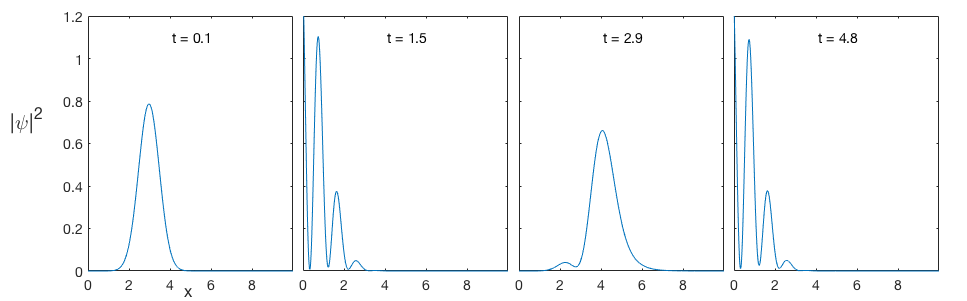}
\caption{Snapshots of $|\psi(x,t)|^2$ with the initial data $f(x) = \frac{e^{-(x-3)^2}}{\sqrt[4]{\pi/2}}$, and parameters $\Lambda = -1$ and $\alpha = 1.0$. The Universe moves toward the origin $(t=0.1 - 1.5)$, expands asymmetrically $(t=2.9)$, and contracts again $(t=4.8)$.  The profiles at $t=1.5$ and $t= 4.8$ are nearly identical. }
\label{fig1}
\end{figure}

 Convoluting  the data so extended with the full-line propagator (\ref{prop}) then gives the solution 
\bea
  \psi(x,\zeta) &=& \int_{-\infty}^0 dx' \ K(x,\zeta;x',0)\ f_L(x') \nn\\
    &&+ \int_0^\infty dx' \ K(x,\zeta;x',0)\ f(x'),   \ \  x>0.
\eea
It is straightforward to construct explicit examples of such solutions; all describe Universes that  expand out to a maximum size, re-collapse, and bounce again. This is of course expected since wave packets are confined in the half-oscillator potential. Figure (\ref{fig1}) shows the dynamics of a representative Gaussian wave function with $\Lambda=-1$, and $\alpha=1$.  The asymmetric bounce is evident, and the second and fourth frames demonstrate the multiple bounce feature.  

\smallskip

\noindent \underbar{$\Lambda >0$:}  The Hamiltonian is not bounded below. However the unitary evolution operator is still well defined  since the Hamiltonian has self-adjoint extensions.  The propagator on $\mathbb{R}$ is obtained by the replacement $l\rightarrow il$ to give
\bea
\bar{K}(x,\zeta; x',0) &=& \sqrt{\frac{1}{2\pi il\sinh{\zeta}}} \nn\\
  &&\times  \exp\left\{ \frac{i[(x^2 +x'^2)\cosh \zeta -2xx' ]}{2l\sinh \zeta}    \right\}.\label{prop2}
\eea
 Solutions of the time-dependent Schrodinger equation with the boundary condition $\phi'(0)-\alpha \phi(0)=0$ are found in the same way as above by extending the initial data function to $x<0$. It is evident that the propagator is damped for large times $\zeta$ due to the prefactor. However for the very small $\Lambda$ that is experimentally observed, the decay time would be very large. (It is useful to note that the issue of convergence of the Euclidean functional integral for the inverted oscillator was studied in  \cite{Carreau:1990is}, where it is shown that the integral for the propagator converges if the propagation time is bounded by a factor of the oscillator frequency.) Fig. 2 shows the propagation of the same initial Gaussian wave packet as that in Fig. 1, but now  for positive $\Lambda$. The wave packet moves outward and spreads rapidly.

 \begin{figure}
\includegraphics[width = \linewidth]{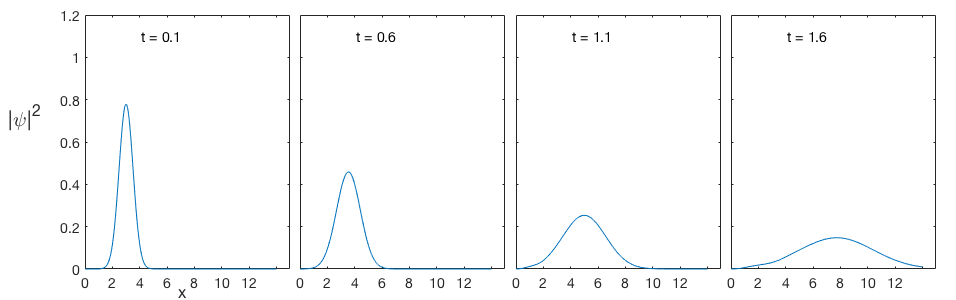}
\caption{Snapshots of $|\psi(x,t)|^2$ with the initial data $f(x) = \frac{e^{-(x-3)^2}}{\sqrt[4]{\pi/2}}$, and parameters $\Lambda = 1$ and $\alpha = 1.0$. The initial  wave packet travels outwards and spreads.  }
\label{fig2}
\end{figure}
 
 \subsection{Quantization on $\mathbb{R}$ }
 
In the above we started with the standard canonical parametrization for the FLRW cosmology which led to the oscillator on the half-line. 
There is an alternative  parametrization that directly gives the oscillator on the real line after a rescaling of variables. This is  
\bea
q_{ab} &=&  A^{4/3}(t) e_{ab} \nn\\
\pi^{ab} &=& \frac{1}{4 A^{1/3}(t)} \ P_A(t) e^{ab},
\eea
where the phase space $(A,P_A)$ is now $\mathbb{R}^2$. 

In this parametrization there is an exact Lorentzian ``Hartle-Hawking'' wave function, which is the  amplitude to create the universe from nothing, albeit in the dust time gauge. This is  obtained from (\ref{prop2}) :
\be
\Psi_{HH} \equiv\bar{K}(A,\zeta; 0,0) = \sqrt{\frac{1}{2\pi i l \sinh{\zeta}}} 
   \exp\left( -\frac{ iA^2}{2l\tanh \zeta}    \right),\label{prop3}
\ee 
where $A^4= \text{det}(q_{ab})\equiv q$, and since we are now on the full line, $A\in \mathbb{R}$. This expression is just the oscillator propagator on the real line for $\Lambda = 1/l^2$ with $A_0=\zeta_0=0$.  
For large times $\zeta=t/l$ this is 
\be
\bar{K}(q,\zeta; 0,0) \longrightarrow    \frac{1}{\sqrt{\pi i l}} \exp\left(-\frac{ i\sqrt{q} + t}{2l }\right) .  \label{prop3larget}  
\ee
This is oscillatory in 3-volume, and decays exponentially in time $t$.  
\section{Discussion}

The basic result in this note is that in general relativity coupled to pressureless dust in the dust time gauge, the FLRW model with a cosmological constant has a physical Hamiltonian that is exactly that of a harmonic oscillator with frequency determined by $\sqrt{\Lambda}$.   The Hamiltonian has a one-parameter ($\alpha$) set of self-adjoint extensions, and explicit solutions of the time-dependent Schrodinger equation are readily constructed. All cases give singularity avoidance, which here means that wave functions describing the Universe bounce at small spatial volume for any value of $\alpha$, regardless of whether the configuration space is the half line or the  full line. 

It is interesting to compare these results with those obtained in LQC \cite{Agullo:2016tjh} using the connection-triad variables. There the $\Lambda =0$ case  was studied with scalar field time, where the form of the Hamiltonian is such that wave function dynamics requires    numerical study.   It was subsequently studied in the dust time \cite{Husain:2011tm}. In both these cases the Hamiltonian  is essentially self-adjoint. In our case the bounce occurs for all self-adjoint extensions, and can be asymmetric in the sense that there is a phase shift at the bounce determined by $\alpha$.  Only the $\alpha=0$ case gives a symmetric bounce.

For comparison with Dirac quantization,  the corresponding quantum theory also resembles the oscillator, but only for the Laplace-Beltrami operator ordering in the kinetic term in the Wheeler-DeWitt operator \cite{Maeda:2015fna};  this work (which was pointed out to us after the present work was posted to the arXiv) considered only $\Lambda=1$, and did not address the most general self-adjoint extension with Robin boundary conditions. Nevertheless, it is one of the few cases where it seems possible to rigorously establish equivalence between Dirac and reduced phase space quantizations.   It would be interesting to study this issue for full quantum gravity with dust time \cite{Husain:2011tk}. 

Our consideration and results are entirely in the Lorentzian theory, and as such may be compared with similar models that invoke the Hartle-Hawking prescription in Lorentzian time, in particular the recent debate concerning integration contours for the propagator \cite{Feldbrugge:2017kzv,DiazDorronsoro:2017hti}. The latter work  reports a suppression factor $\exp(-\Lambda l_p^2)$ in the propagator  for the no boundary wave function of the Universe in the semiclassical approximation. We find a similar result, but our state is {\it exact}, (i.e. not just a semiclassical approximation), and also has explicit (dust) time dependence: eqn. (\ref{prop3larget}) has a suppression factor $\exp(-t/2l)$. From the currently observed value of $\Lambda$, $ l \sim 10^{60} l_p$, therefore the characteristic  decay time is  $\sim 10^{60}$ Planck times, which is close to the age of the Universe. 

The model with spatial curvature $k\ne 0$  and additional matter fields such as the minimally coupled scalar field is not exactly solvable. The physical Hamiltonian for this case in the dust time gauge (after the canonical transformation (\ref{canon}) ) is 
\be
\Hp^k =\frac{1}{2}\left({p^2} - \Lambda x^2 \right) + k x^{2/3} + \frac{p_\phi^2}{2 x^2} + x^2 V(\phi).  
\ee 
Gravitational perturbations can be added in a similar way. Models such as this demonstrate that it is useful to consider matter time gauges in the cosmological setting.  

 Lastly the $\Lambda<0$ case may be of interest in the context of the AdS/CFT conjecture and holography. Specifically the idea of  using matter (or  other) time gauge in the bulk might provide a useful mechanism to probe bulk dynamics and the holographic signatures of resolved singularities in such settings \cite{Bodendorfer:2018dgg}, something which appears so far to be largely unexplored.

   \medskip
   \noindent {\bf Acknowledgements}  This work was supported by the Natural Science and Engineering Research Council of Canada. This work was initiated while V. H. was visiting the Perimeter Institute. Research at Perimeter Institute is supported  by  the  Government  of  Canada  through  Industry  Canada  and  by  the  Province  of
Ontario through the Ministry of Research and Innovation.

\bibliography{wavefunction}
\end{document}